# LATTICE RESULTS FOR HEAVY QUARK PHYSICS

H. WITTIG (UKQCD COLLABORATION)

*Physics Department, The University, Southampton SO17 1BJ, UK*

The status of lattice calculations for heavy quark systems is reviewed, focussing on weak matrix elements for leptonic and semi-leptonic decays of heavy mesons. After an assessment of the main systematic errors, results for the decay constants $f_D$ and $f_B$, the $B$ parameter describing $B-\bar{B}$ mixing, the Isgur-Wise function and the spectroscopy of heavy baryons are discussed.

## 1  Introduction

The physics of hadrons containing heavy quarks plays an important rôle in the study of some of the CKM matrix elements, and therefore serves to test the consistency of the Standard Model. Lattice simulations of QCD provide a non-perturbative treatment of hadronic processes, and are thus capable of dealing with large strong interaction effects in weak decay amplitudes.

Lattice QCD replaces space-time by a four-dimensional hypercubic lattice of size $L^3 \cdot T$. The sites are separated by the lattice spacing $a$, which acts as an UV cut-off. One problem encountered in current simulations is that typical values of $a^{-1}$ lie in the range $2-3.5\,\mathrm{GeV}$. Therefore one expects that discretisation errors ("lattice artefacts") will distort the results already for charm physics. Also, $b$ quarks cannot be studied directly, since their mass is above the UV cut-off. One way of addressing this problem is to subtract the leading discretisation error by employing an $O(a)$-improved lattice action[1]. Quantities can then be computed safely around $m_{\mathrm{charm}}$ and extrapolated to the $b$ quark mass. Alternatively, one can use the "static approximation" and perform the simulation at infinite heavy quark mass, using the leading term of an expansion of the heavy quark propagator in $1/m_Q$. Furthermore, $b$ quarks can be formulated using a non-relativistic expression for the QCD action. These three methods provide complementary information for $b$ quark physics on the lattice.

The other main systematic errors which affect our results include using the Quenched Approximation, in which the effects of quark loops are neglected. Furthermore, matrix elements computed on the lattice are related to their continuum counterparts via finite renormalisation constants, due to explicit chiral symmetry breaking induced by the fermionic lattice action. The numerical values of these matching factors are subject to large uncertainties. Finally, lattice estimates of dimensionful quantities are affected by uncertainties in the lattice scale, which arise from the fact that different quantities used to estimate

Table 1: Lattice estimates for the pseudoscalar decay constants from different collaborations

| Collab. | $f_B$ [MeV] | $f_D$ [MeV] | $f_{B_s}/f_B$ | $f_{D_s}/f_D$ |
|---|---|---|---|---|
| MILC[3] | 148(5)(14)(19) | 180(4)(18)(16) | 1.13(2)(9) | 1.09(1)(4) |
| UKQCD[4] | 160 $^{+6}_{-6}$ $^{+53}_{-19}$ | 185 $^{+4}_{-3}$ $^{+42}_{-7}$ | 1.22 $^{+4}_{-3}$ | 1.18 $^{+2}_{-2}$ |
| PCW[5] | 180(50) | 170(30) | 1.09(2)(5) | 1.09(2)(5) |
| BLS[6] | 187(10)(34)(15) | 208(9)(35)(12) | 1.11(6) | 1.11(6) |
| ELC[7] | 205(40) | 210(15) | 1.08(6) | 1.10(4) |

$a^{-1}$ [GeV] give different results.

Most of the results discussed here come from a simulation by the UKQCD Collaboration, using a lattice of size $24^3 \cdot 48$ at $\beta = 6/g^2 = 6.2$ for which $a^{-1} = 2.9(2)$ GeV.[2] For propagating quarks an $O(a)$-improved fermion action was used. Results for $b$ physics were obtained using propagating heavy quarks and also the static appoximation.

## 2 Leptonic $B$ decays and $B^0 - \overline{B^0}$ mixing

The decay constant of a heavy-light pseudoscalar meson, $f_P$, can be extracted from the matrix elements of the lattice axial current via

$$\langle 0|A_4^{\text{latt}}(0)|P\rangle \sim M_P f_P/Z_A \qquad (1)$$

where $M_P$ is the pseudoscalar mass, and $Z_A = O(1)$ is the matching factor relating the lattice matrix element to its continuum counterpart. In Fig. 1 we show an example for the scaling behaviour of $f_P\sqrt{M_P}$ with increasing heavy quark mass. According to HQET, this quantity behaves like a constant, but, as Fig. 1 shows, there are large deviations from this scaling law[4]. The estimates for $f_B$, $f_D$ and the ratios $f_{B_s}/f_B$, $f_{D_s}/f_D$ from various collaborations are shown in Tab. 1. A weighted average of the results in the table yields

$$f_B = 168 \pm 30\,\text{MeV}, \qquad f_D = 196 \pm 20\,\text{MeV} \qquad (90\%\,\text{C.L.}) \qquad (2)$$

The phenomenologically interesting quantity for the study of CKM matrix elements is the combination $f_B\sqrt{B_B}$, which requires knowledge of the $B$ parameter $B_B$, defined via $B_B = \alpha_s(\mu)^{-2/\beta_0} \langle \overline{B^0}|O_L(\mu)|B^0\rangle/\frac{8}{3} f_B^2 M_B^2$, where $O_L$ is the $\Delta B = 2$ four-fermion operator. In a recent study, UKQCD presented results for $B_B$ and $f_B$ using the static approximation[2]. The results for the $B$ parameter are

$$B_{B_d} = 0.99\ ^{+5}_{-6}\ ^{+3}_{-2}, \qquad B_{B_s} = 1.01\ ^{+4}_{-5}\ ^{+2}_{-1}, \qquad (3)$$

Figure 1: UKQCD's data for the quantity $f_P\sqrt{M_P}\,\alpha_s(M_P)^{2/\beta_0}$ plotted versus $1/M_P$ in lattice units. The solid line denotes the extrapolation of the four points using propagating quarks. The point at infinite meson mass is obtained using the static approximation.

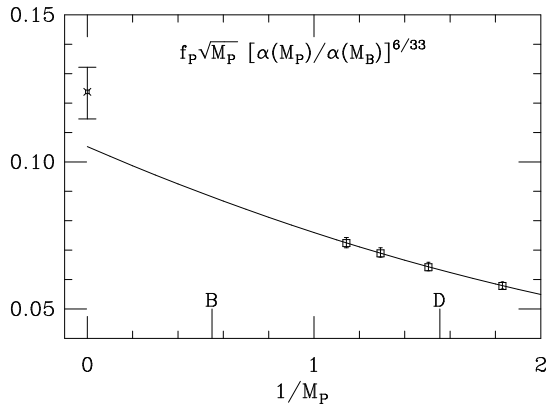

where the first error is statistical and the second systematic. It should be stressed that there is a further 15–20% uncertainty coming from the matching procedure between the lattice and continuum matrix elements. It is reasonable, however, to expect that systematic effects cancel partially in the dimensionless ratios[2]

$$\frac{f_{B_s}\sqrt{B_{B_s}}}{f_{B_d}\sqrt{B_{B_d}}} = 1.16 \;{}^{+\,4}_{-\,4}\,(\text{stat})\;{}^{+\,2}_{-\,1}\,(\text{syst}) \qquad (4)$$

$$\frac{f^2_{B_s} B_{B_s} M_{B_s}}{f^2_{B_d} B_{B_d} M_{B_d}} = 1.37 \;{}^{+\,9}_{-\,8}\,(\text{stat})\;{}^{+\,5}_{-\,4}\,(\text{syst}), \qquad (5)$$

These results can be applied to the ratio of $B^0 - \overline{B^0}$ mixing parameters $x_s/x_d$ given by

$$\frac{x_s}{x_d} = \frac{\tau_{B_s}}{\tau_{B_d}}\,\frac{\hat\eta_{B_s}}{\hat\eta_{B_d}}\,\frac{M_{B_s}}{M_{B_d}}\,\frac{f^2_{B_s} B_{B_s}}{f^2_{B_d} B_{B_d}}\,\frac{|V_{ts}|^2}{|V_{td}|^2}. \qquad (6)$$

where $\tau_{B_{d,s}}$ denote the $B_d$ and $B_s$ meson lifetimes, and $\hat\eta_{B_{d,s}}$ parametrise short-distance QCD corrections. Assuming $\hat\eta_{B_d} = \hat\eta_{B_s}$ and using $\tau_{B_d} = 1.53(9)\,\text{ps}$, $\tau_{B_s} = 1.54(14)\,\text{ps}$, as well as our lattice estimate eq. (5), we find

$$x_s/x_d = (1.38 \pm 0.17)\,|V_{ts}|^2/|V_{td}|^2. \qquad (7)$$

In conjunction with the experimental value $x_d = 0.76(6)$[8], the above ratios can also also be used to predict $x_s$, provided $\frac{|V_{ts}|^2}{|V_{td}|^2}$ is constrained using global fits[9].

For a *fixed* value of the $B$ parameter $B_K$, $x_s$ is then obtained as a function of $f_B\sqrt{B_B}$. Choosing $B_K = 0.8$, we find $x_s \simeq 14$ for $f_B\sqrt{B_B} \simeq 170\,\text{MeV}^2$.

## 3 Semi-leptonic $B$ decays and the Isgur-Wise function

Processes like $B \to D\ell\nu_\ell$ and $B \to D^*\ell\nu_\ell$ can be parametrised in terms of six form factors, e.g.

$$\frac{\langle D|V_\mu|B\rangle}{\sqrt{M_B M_D}} = (v_B + v_D)_\mu h_+(\omega) + (v_B - v_D)_\mu h_-(\omega) \tag{8}$$

where $\omega = v_B \cdot v_D$ denotes the product of 4-velocities if the $B$ and $D$ mesons. The Heavy Quark Symmetry (HQS) relates the six form factors to one universal form factor, $\xi(\omega)$, called the Isgur-Wise function, which is normalised at zero recoil, $\xi(\omega = 1) = 1$. The form factor $h_+(\omega)$ is related to $\xi(\omega)$ via

$$h_+(\omega) = (1 + \beta_+(\omega) + \gamma_+(\omega))\, \xi(\omega) \tag{9}$$

where $\beta_+(\omega)$ parametrises radiative corrections between HQET and full QCD, and $\gamma_+(\omega)$ denotes (unknown) corrections in $1/m_Q$.

In our lattice study we concentrate on pseudoscalar ($P$) to pseudoscalar ($P'$) transitions and extract the lattice form factor $h_+(\omega)^{\text{latt}}$ from the matrix element of the temporal component of the vector current. We determine the Isgur-Wise function through the ratio

$$R_+(\omega) \equiv \frac{h_+(\omega)^{\text{latt}}\,(1 + \beta_+(1))}{h_+(1)^{\text{latt}}\,(1 + \beta_+(\omega))} \simeq \xi(\omega) \tag{10}$$

which, as is argued in [10,11], is a good approximation, provided that the $\omega$-dependence of lattice artefacts and in $\gamma_+(\omega)$ is weak.

In UKQCD's lattice simulation using propagating heavy quarks, one observes a weak dependence of $R_+(\omega)$ on the heavy quark mass. Furthermore, repeating the analysis for the axial form factor $h_{A_1}(\omega)$ relevant for $B \to D^*\ell\nu_\ell$ decays, gives $R_+(\omega)/R_{A_1}(\omega) \simeq 1$[11]. These observations support the universality of $\xi(\omega)$: one obtains the same estimate for the Isgur-Wise function over a range of different heavy quark masses (i.e. for different flavours), and also if the spin structure of the outgoing meson is changed.

The lattice data for $\xi(\omega)$ are obtained for light quark masses around the strange quark mass and can be extrapolated to the chiral limit. The extrapolated Isgur-Wise function can then be parametrised in the form

$$\xi_{\text{NR}}(\omega) = \tfrac{2}{\omega+1} \exp\left\{-(2\rho^2 - 1)\tfrac{\omega-1}{\omega+1}\right\}, \tag{11}$$

Table 2: Lattice results for the slope of the Isgur-Wise function at zero recoil.

| Collab. | $\rho_s^2$ | $\rho_{u,d}^2$ |
|---|---|---|
| UKQCD[10] | $1.2\ ^{+2}_{-2}\ ^{+2}_{-1}$ | $0.9\ ^{+2}_{-3}\ ^{+4}_{-2}$ |
| BSS[12] | $1.24(26)(36)$ | |
| MO[13] | $0.95$ | |

Figure 2: UKQCD's estimate of $|V_{cb}|\xi_{\mathrm{NR}}(\omega)$ compared to the CLEO data.

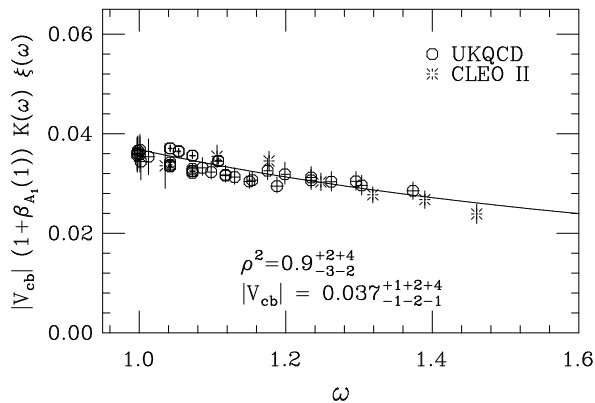

where $\rho^2$ is the slope of $\xi(\omega)$ at zero recoil. Lattice results for $\rho^2$ for light quark masses either at the strange quark mass or in the chiral limit from various collaborations are shown in Tab. 2.

In order to extract $|V_{cb}|$ we can now use the parametrised form of $\xi(\omega)$ to extrapolate the measured decay rate for $B \to D^*\ell\nu_\ell$ decays to $\omega = 1$. Fitting the CLEO II data[14] to our estimate of $|V_{cb}|\xi_{\mathrm{NR}}$ (neglecting radiative and power corrections), we obtain

$$|V_{cb}| = 0.037\ ^{+1}_{-1}(\mathrm{exp})\ ^{+2}_{-2}(\mathrm{stat})\ ^{+4}_{-1}(\mathrm{syst}). \qquad (12)$$

A comparison of the experimental data and our fit is shown in Fig. 2.

## 4  Heavy Baryon Spectroscopy

Lattice studies can help making predictions for the masses of $\Lambda$, $\Sigma$, $\Xi$ and $\Omega$ states in both the charm and beauty sector. Baryon masses are extracted from the exponential fall-off of correlation functions of suitably chosen interpolating operators of baryonic states.

Table 3: Heavy baryon states and their masses measured experimentally and on the lattice[15].

| Baryon $h = c, b$ | $J^P$ | $(I)\ (S)\ s_l^{\pi_l}$ | Mass [GeV] | Quark Content | Lattice [GeV] |
|---|---|---|---|---|---|
| $\Lambda_h$ | $\frac{1}{2}^+$ | $(0)(0)\ 0^+$ | 2.285(1) | $(ud)c$ | $2.28\ ^{+\ 4}_{-\ 4}$ |
|  |  |  | 5.64(5) | $(ud)b$ | $5.59\ ^{+\ 9}_{-10}$ |
| $\Sigma_h$ | $\frac{1}{2}^+$ | $(1)(0)\ 1^+$ | 2.453(1) | $(uu)c$ | $2.45\ ^{+\ 6}_{-\ 4}$ |
|  |  |  |  | $(uu)b$ | $5.69\ ^{+10}_{-10}$ |
| $\Sigma_h^*$ | $\frac{3}{2}^+$ | $(1)(0)\ 1^+$ | 2.53(5)(?) | $(uu)c$ | $2.43\ ^{+\ 5}_{-\ 4}$ |
|  |  |  |  | $(uu)b$ | $5.68\ ^{+\ 9}_{-10}$ |
| $\Xi_h$ | $\frac{1}{2}^+$ | $(\frac{1}{2})(-1)\ 0^+$ | 2.470(4) | $(us)c$ | $2.41\ ^{+\ 3}_{-\ 3}$ |
|  |  |  |  | $(us)b$ | $5.69\ ^{+\ 7}_{-10}$ |
| $\Omega_h$ | $\frac{1}{2}^+$ | $(0)(-2)\ 1^+$ | 2.74(2) | $(ss)c$ | $2.68\ ^{+\ 4}_{-\ 4}$ |
|  |  |  |  | $(ss)b$ | $5.91\ ^{+\ 7}_{-\ 8}$ |

The results by UKQCD[15] are shown in Tab. 3 together with the experimental numbers as far as these are known. There is good agreement between lattice and experiment for charmed baryons, and also the lattice value of the mass of the $\Lambda_b$ is consistent. Taking the uncertainty in the lattice scale $a^{-1}$ into account, UKQCD's result is $M_{\Lambda_b} = 5.59\ ^{+\ 9}_{-10}(\text{stat})\ ^{+39}_{-39}(\text{syst})\,\text{GeV}$, whereas the PSI-CERN-Wuppertal Collaboration[16] obtain $M_{\Lambda_b} = 5.728 \pm 0.144 \pm 0.018\,\text{GeV}$. Both these numbers are consistent with the experimental value of $M_{\Lambda_b} = 5.64 \pm 0.05\,\text{GeV}$[17].

Mass splittings between the $\Lambda$, $\Sigma$ and/or pseudoscalar mesons ($P$) are also obtained. Tab. 4 shows UKQCD's data for the dimensionless splittings $R_{\Lambda_{c,b}}$ and $R_{\Sigma_{c,b}}$, which show excellent agreement with the experimental numbers. The mass of the $\Sigma^*$ seems to be slightly below that of the $\Sigma$, as can be seen from Tab. 4. In fact, our lattice estimates for the $\Sigma$ and $\Sigma^*$ masses are statistically compatible. Recent measurements[18], however, indicate that the $\Sigma^* - \Sigma$ splitting may be so small that it cannot be resolved with our present statistics.

In future we will study semi-leptonic $\Lambda_b \to \Lambda_c \ell \nu_\ell$ decays. Preliminary results are presented elsewhere[19].

Table 4: UKQCD's results for the mass splittings $R_{\Lambda_h} \equiv (M_{\Lambda_h} - M_{P_h})/(M_{\Lambda_h} + M_{P_h})$ and $R_{\Sigma_h} \equiv (M_{\Sigma_h} - M_{\Lambda_h})/(M_{\Sigma_h} + M_{\Lambda_h})$, $h = c, b$ compared to experiment.

|  | $R_{\Lambda_c}$ | $R_{\Lambda_b}$ | $R_{\Sigma_c}$ | $R_{\Sigma_c}$ |
|---|---|---|---|---|
| Lattice | $0.099\,^{+9}_{-7}$ | $0.033\,^{+5}_{-4}$ | $0.038\,^{+9}_{-9}$ | $0.017\,^{+5}_{-7}$ |
| Exp. | 0.100(3) | 0.033(5) | 0.035(1) | - |

**Acknowledgments**

I would like to thank the organisers of this conference for creating such a stimulating environment. This work was supported by the Particle Physics and Astronomy Research Council. I thank my colleagues from the UKQCD Collaboration for many fruitful discussions.